\documentclass[12pt]{article}
\usepackage{graphicx}
\usepackage{amsmath}
\usepackage{amssymb}
\usepackage{caption2}
\begin{document}
\bibliographystyle{prsty}
\begin{center}
{\large {\bf \sc{   Analysis of the vertices $\rho
NN$, $\rho\Sigma\Sigma$ and $\rho\Xi\Xi$ with light-cone QCD sum rules  }}} \\[2mm]
Zhi-Gang Wang \footnote{E-mail,wangzgyiti@yahoo.com.cn.  }    \\
Department of Physics, North China Electric Power University,
Baoding 071003, P. R. China
\end{center}

\begin{abstract}
In this article, we calculate the strong coupling constants of the
$\rho NN$, $\rho\Sigma\Sigma$ and $\rho\Xi\Xi$ in the framework of
the light-cone QCD sum rules approach. The  strong coupling
constants of the meson-baryon-baryon are the fundamental parameters
in the one-boson exchange model which describes the baryon-baryon
interactions successfully. The numerical values are in agreement
with the existing calculations in part. The electric and magnetic
$F/(F+D)$ ratios deviate  from the prediction of the vector meson
dominance theory, the $SU(3)$ symmetry breaking effects are very
large.
\end{abstract}


\section{Introduction}

One of the  successful approaches in describing the two-baryon
interactions is the one-boson exchange (or the Nijmegen soft-core
potential) model \cite{Nijmegen,Nijmegen2}. In this model, the
baryon-baryon interactions are mediated by the  intermediate mesons,
such as the pseudoscalar octet mesons $\pi$, $K$, $\eta$, the vector
octet mesons $\rho$, $K^*$, $\omega$ and the scalar octet mesons
$\sigma$, $a_0$, $f_0$, etc. The  strong coupling constants of the
meson-baryon-baryon are the fundamental parameters, they have been
empirically determined (or fitted)  to reproduce the data of the
nucleon-nucleon, hyperon-nucleon  and  hyperon-hyperon interactions.
The strong coupling constants of the vector mesons with the octet
baryons (thereafter we will denote them by $g_{VNN}$) can be written
in term of the $\rho NN$ couplings  and the electric (and magnetic)
$F/(F+D)$ ratios. The vector meson dominance theory indicates that
the electric $F/(F+D)$ ratio $\alpha_e$ be $\alpha_e=1$ via the
universal coupling of the $\rho$ meson to the isospin current
\cite{VMD}.

It is important to determine those fundamental quantities directly
from the quantum chromodynamics. Based on the assumption of the
 strong couplings between the quarks and vector mesons, the $g_{VNN}$
 have been calculated in the external field QCD sum rules    approach, while
the  coupling constants of the quark-meson were determined in some
phenomenological models \cite{Rijken06}. The strong coupling
constants of the  scalar mesons with the octet baryons  have also
been calculated in the external field QCD sum rules \cite{Oka06}. In
the external field QCD sum rules, the operator product expansion is
used to expand the time-ordered currents into a series of quark
condensates, gluon condensates and vacuum susceptibilities which
parameterize the
 long distance properties of  the QCD vacuum and the
non-perturbative interactions of the quarks and gluons with the
external field \cite{EQCDSR}.

 In this article, we calculate the strong coupling constants of the $\rho NN$,
 $\rho \Sigma\Sigma$ and $\rho\Xi\Xi$ in the framework of the
 light-cone QCD sum rules approach, and determine the electric (and
 magnetic) $F/(F+D)$ ratios $\alpha_e$ ( and $\alpha_m$). The strong
 coupling constants of the $\rho NN$ have been calculated with the
 light-cone QCD sum rules approach \cite{Zhu98}, we revisit this subject and obtain
 different predictions. Furthermore, the strong coupling constants of the
 pseudoscalar mesons with the octet baryons  have also been calculated with the
 light-cone QCD sum rules \cite{Aliev06}.
 The light-cone QCD sum
rules approach carries out the operator product expansion near the
light-cone $x^2\approx 0$ instead of the short distance $x\approx 0$
while the non-perturbative hadronic matrix elements  are
parameterized by the light-cone distribution amplitudes
 which classified according to their twists  instead of
 the vacuum condensates \cite{LCSR,LCSRreview}. The non-perturbative
 parameters in the light-cone distribution amplitudes are calculated by   the conventional QCD  sum rules
 and the  values are universal \cite{SVZ79}.

The article is arranged as: in Section 2, we derive the strong
coupling constants  $\rho NN$, $\rho\Sigma\Sigma$ and $\rho\Xi\Xi$
in the light-cone QCD sum rules approach; in Section 3, the
numerical result and discussion; and in Section 4, conclusion.

\section{Strong coupling constants  $\rho NN$, $\rho\Sigma\Sigma$ and $\rho\Xi\Xi$ with light-cone QCD sum rules}

In the following, we write down the
 two-point correlation functions $\Pi_i(p,q)$,
\begin{eqnarray}
\Pi_N(p,q)&=&i \int d^4x \, e^{-i q \cdot x} \,
\langle 0 |T\left\{ J_N(0)\bar{J}_N(x)\right\}|\rho(p)\rangle \, , \\
\Pi_\Sigma(p,q)&=&i \int d^4x \, e^{-i q \cdot x} \, \langle 0
|T\left\{ J_\Sigma(0)\bar{J}_\Sigma(x)\right\}|\rho(p)\rangle \, ,
\\
\Pi_\Xi(p,q)&=&i \int d^4x \, e^{-i q \cdot x} \,
\langle 0 |T\left\{ J_\Xi(0)\bar{J}_\Xi(x)\right\}|\rho(p)\rangle \, , \\
J_N(x)&=& \epsilon^{abc} u^T_a(x)C\gamma_\mu u_b(x) \gamma_5 \gamma^\mu d_c(x)\, ,  \nonumber \\
J_\Sigma(x)&=& \epsilon^{abc} u^T_a(x)C\gamma_\mu u_b(x) \gamma_5 \gamma^\mu s_c(x)\, ,  \nonumber \\
J_\Xi(x)&=& \epsilon^{abc} s^T_a(x)C\gamma_\mu s_b(x) \gamma_5
\gamma^\mu u_c(x)\, ,
\end{eqnarray}
where the baryon currents $J_N(x)$, $J_\Sigma(x)$ and $J_\Xi(x)$
 interpolate the octet baryons $p$, $\Sigma$ and $\Xi$,
 respectively \cite{Ioffe81}, the external  state $\rho_0$ has the
four momentum $p_\mu$ with $p^2=m_\rho^2$ .

The vector meson $V_\mu$ can couple  with the vector current $J_\mu$
with the following Lagrangian,
\begin{eqnarray}
 \mathcal {L}&=&-gJ_\mu (x) V^\mu(x) \, ,
 \end{eqnarray}
where the $g$ denotes  the coupling constant. The form factors of
the vector current between two octet baryons can be written as
\begin{eqnarray}
\langle N(p_1)| J_\mu (0)|N(p_2)\rangle&=& \overline{N}(p_1)
\left\{g_V\gamma_\mu +i
g_T\frac{\sigma_{\mu\nu}p^\nu}{2m}+g_P\frac{p_\mu}{2m}
\right\}N(p_2) \, , \nonumber\\
\int dx\langle N(p_1)|  i\mathcal {L}(x)|N(p_2) \rho(p)\rangle&=&
-i\overline{N}(p_1) \left\{g_V\!\not\!{\epsilon} +i
g_T\frac{\epsilon^\mu\sigma_{\mu\nu}p^\nu}{2m} +g_P\frac{p \cdot
\epsilon}{2m}\right\}N(p_2) \, , \nonumber\\
\end{eqnarray}
where the $m$ is the average value of the masses of the  two octet
baryons. In the limit $p^2=m_\rho^2$, $\epsilon \cdot p=0$, the form
factors $g_V(p^2=m_\rho^2)$ and $g_T(p^2=m_\rho^2)$ are reduced to
the strong coupling constants of the phenomenological Lagrangian,
\begin{eqnarray}
\mathcal {L}&=&-g_V \bar{\psi}\gamma_\mu \psi
V^\mu+\frac{g_T}{4m}\bar{\psi}\sigma^{\mu \nu} \psi (\partial_\mu
V_\nu-\partial_\nu V_\mu) \, .
\end{eqnarray}

According to the basic assumption of current-hadron duality in the
QCD sum rules approach \cite{SVZ79}, we can insert  a complete
series of intermediate states with the same quantum numbers as the
current operators $J_N(x)$, $J_\Sigma(x)$    and $J _\Xi(x)$ into
the correlation functions $\Pi_{i}(p,q)$  to obtain the hadronic
representation. After isolating the ground state contributions from
the pole terms of the baryons $p$, $\Sigma$    and $\Xi$, we get the
following results,
\begin{eqnarray}
\Pi_{i }(p,q)&=&
\frac{\lambda_i^2}{\left[m_i^2-(q+p)^2\right]\left[m_i^2-q^2\right]}
\left\{-\left[g_V+g_T\right]\frac{m_\rho^2}{2}\!\not\!{\epsilon}
-g_V \epsilon \cdot q \left[2\!\not\!{q}
+\!\not\!{p}\right]\right\}+\cdots \, , \nonumber \\
&=&\Pi_i^1(p,q)\!\not\!{\epsilon}+\Pi_i^2(p,q)\epsilon \cdot q
\!\not\!{q} +\cdots \, ,
\end{eqnarray}
where the following definitions have been used,
\begin{eqnarray}
\langle 0| J_i (0)|N(p)\rangle &=&\lambda_i U(p) \, , \nonumber \\
U(p)\overline {U}(p)&=&\!\not\!{p}+m_N \, ,
\end{eqnarray}
here we use the notation $N$ to represent the octet baryons $p$,
$\Sigma$ and $\Xi$. We have not  shown the contributions from the
single pole terms in Eq.(8) explicitly, they can be deleted
completely after the double Borel transformation. In the original
QCD sum rules analysis of the nucleon magnetic
 moments \cite{EQCDSR}, the interval of dimensions (of the condensates) for the chiral odd
structures is larger than the interval of dimensions for the chiral
even structures, one may expect a better accuracy of the results
obtained from the sum rules with  the chiral odd structures.
 In this article, we choose the tensor structures
$\!\not\!{\epsilon}$ and $\epsilon \cdot q \!\not\!{q}$ for
analysis.

In the following, we briefly outline the  operator product expansion
for the correlation functions $\Pi_{i }(p,q)$  in perturbative QCD
theory. The calculations are performed at the large space-like
momentum regions $(q+p)^2\ll 0$  and $q^2\ll 0$, which correspond to
the small light-cone distance $x^2\approx 0$ required by the
validity of the operator product expansion approach. We write down
the "full" propagator of a massive light  quark in the presence of
the quark and gluon condensates firstly \cite{SVZ79}\footnote{One
can consult the second article of Ref.\cite{SVZ79} for the technical
details in deriving the full propagator.},
\begin{eqnarray}
\langle0 | T[q_a(x)\overline{q}_b(0)]|0\rangle &=&
\frac{i\delta_{ab}\!\not\!{x}}{ 2\pi^2x^4}
-\frac{\delta_{ab}m_q}{4\pi^2x^2}-\frac{\delta_{ab}}{12}\langle
\bar{q}q\rangle +\frac{i\delta_{ab}}{48}m_q
\langle\bar{q}q\rangle-\frac{\delta_{ab}x^2}{192}\langle
\bar{q}g_s\sigma Gq\rangle
\nonumber\\
&& +\frac{i\delta_{ab}x^2}{1152}m_q\langle \bar{q}g_s\sigma
Gq\rangle \!\not\!{x}-\frac{i}{32\pi^2x^2}\frac{\lambda^A_{ab}}{2}
G^A_{\mu\nu} (\!\not\!{x} \sigma^{\mu\nu}+\sigma^{\mu\nu}
\!\not\!{x})  \nonumber\\
&&+\cdots \, ,
\end{eqnarray}
then contract the quark fields in the correlation functions
$\Pi_i(p,q)$ with the Wick theorem, and obtain the results
\begin{eqnarray}
\Pi_N(p,q)&=&2i\epsilon_{abc}\epsilon_{ijk} \int d^4x \, e^{-i q
\cdot x} \, Tr\left[ \gamma_\nu C U^T_{bi}(-x)C\gamma_\mu U_{aj}(-x)\right] \nonumber \\
&&\gamma_5\gamma^\mu D_{ck}(-x)\gamma^\nu \gamma_5\, , \\
\Pi_\Sigma(p,q)&=&2i\epsilon_{abc}\epsilon_{ijk} \int d^4x \, e^{-i
q\cdot x} \, Tr\left[ \gamma_\nu C U^T_{bi}(-x)C\gamma_\mu U_{aj}(-x)\right] \nonumber \\
&&\gamma_5\gamma^\mu S_{ck}(-x)\gamma^\nu \gamma_5\, , \\
\Pi_\Xi(p,q)&=&2i\epsilon_{abc}\epsilon_{ijk} \int d^4x \, e^{-i q
\cdot x} \, Tr\left[ \gamma_\nu C S^T_{bi}(-x)C\gamma_\mu S_{aj}(-x)\right] \nonumber \\
&&\gamma_5\gamma^\mu U_{ck}(-x)\gamma^\nu \gamma_5\, ,
\end{eqnarray}
where the $U$, $D$ and $S$ stand for the full propagators of the
$u$, $d$ and $s$ quarks respectively.
 Take the replacement
\begin{eqnarray}
U^{ab}_{\alpha\beta}(-x)&\rightarrow&\langle 0|u^a_\alpha(0)
\bar{u}^b_\beta(x)|\rho(p)\rangle \, , \\
D^{ab}_{\alpha\beta}(-x)&\rightarrow&\langle 0|d^a_\alpha(0)
\bar{d}^b_\beta(x)|\rho(p)\rangle \, ,
\end{eqnarray}
for one of the quark propagators in the correlation functions in
Eqs.(11-13), perform the following Fierz re-ordering
\begin{eqnarray}
q^a_\alpha(0) \bar{q}^b_\beta(x)&=&-\frac{1}{12}
\delta_{ab}\delta_{\alpha\beta}\bar{q}(x)q(0)
-\frac{1}{12}\delta_{ab}(\gamma^\mu)_{\alpha\beta}\bar{q}(x)\gamma_\mu
q(0) \nonumber\\
&&-\frac{1}{24}\delta_{ab}(\sigma^{\mu\nu})_{\alpha\beta}\bar{q}(x)\sigma_{\mu\nu}q(0) \nonumber\\
&&+\frac{1}{12}\delta_{ab}(\gamma^\mu
\gamma_5)_{\alpha\beta}\bar{q}(x)\gamma_\mu \gamma_5 q(0)
+\frac{1}{12}\delta_{ab}(i \gamma_5)_{\alpha\beta}\bar{q}(x)i
\gamma_5 q(0) \, ,
\end{eqnarray}
 and substitute the hadronic matrix elements ( such as the $ \langle
0| {\bar u} (x) \gamma_\mu u(0) |\rho(p)\rangle$,  $ \langle 0|
{\bar u} (x)  u(0) |\rho(p)\rangle$, $ \langle 0| {\bar u} (x)
\sigma_{\mu\nu} u(0) |\rho(p)\rangle$, etc. )  with
 the corresponding light-cone distribution amplitudes of the $\rho$ meson, finally we
obtain the spectral densities at the coordinate space.

 In calculation, we can encounter some terms like $x^\mu \langle
0| {\bar u} (x) \gamma_\mu d(0) |\rho(p)\rangle$ \footnote{Here we
use the $\rho^-$ meson to illustrate the calculation. }, there are
two approaches to deal with them,
\begin{eqnarray}
  \langle 0| {\bar u} (x) \!\not\!{x} d(0)
|\rho(p)\rangle& =&  f_\rho m_\rho \epsilon \cdot x \int_0^1 du
e^{-i u p\cdot x} \left\{\phi_{\parallel}(u)+\frac{m_\rho^2x^2}{16}
A(u)\right\}\nonumber\\
&&+\frac{1}{2}x^2 \epsilon \cdot x f_\rho m_\rho^3 \int_0^1 du
e^{-iup \cdot x}\int^u_0 dt \int_0^t d\lambda C(\lambda) \, ,
 \end{eqnarray}
and
\begin{eqnarray}
\langle 0| {\bar u} (x)\!\not\!{x} d(0) |\rho(p)\rangle& =&
 if_\rho m_\rho \epsilon \cdot x p \cdot x\int_0^1 du e^{-i u p\cdot x}
\int^u_0 dt \nonumber \\
&&\left\{\phi_{\parallel}(t)-g_{\perp}^{(v)}(t)+\frac{m_\rho^2x^2}{16}
A(t)\right\}\nonumber\\
&&+ \epsilon \cdot x f_\rho m_\rho
\int_0^1 du  e^{-i u p \cdot x} g_{\perp}^{(v)}(u)  \nonumber\\
&&+\frac{1}{2}x^2 \epsilon \cdot x f_\rho m_\rho^3 \int_0^1 du
e^{-iup \cdot x}\int^u_0 dt \int_0^t d\lambda C(\lambda) \, .
 \end{eqnarray}
The two approaches can lead to different results, the analytical
expressions of the second approach are more cumbersome, after the
double Borel transformation and some technical details, we can prove
that the two approaches are equal. In the following, we will present
the analytical results of the first approach only. Once the spectral
densities in the coordinate space are obtained,
 we can translate them  to the
 momentum space with the $D=4+2\epsilon$ dimensional Fourier transformation,

\begin{eqnarray}
\sqrt{2}\Pi_N^1&=&-\frac{1}{2\pi^2}f_\rho m_\rho \int_0^1 du
g_{\perp}^{(v)}(u)\frac{\Gamma(\epsilon-1)}{(-Q^2)^{\epsilon-1}}
\nonumber \\
&&+\frac{1}{2\pi^2}f_\rho m_\rho^3 \int_0^1 du \int^u_0 dt \int_0^t
d\lambda C(\lambda)\frac{\Gamma(\epsilon)}{(-Q^2)^{\epsilon}}
\nonumber\\
&& +\frac{2}{3}\langle \bar{q}q\rangle f^T_\rho m_\rho^2 \int_0^1 du
h_{\parallel}^{(s)}(u)\frac{\Gamma(1)}{(-Q^2)^{1}}\nonumber \\
&&-\frac{1}{12}\langle\frac{ \alpha_sGG}{\pi}\rangle f_\rho
m_\rho \int_0^1 du g_{\perp}^{(v)}(u) \frac{\Gamma(1)}{(-Q^2)^{1}}\nonumber \\
&&-\frac{1}{6}\langle \bar{q}g_s\sigma Gq\rangle f^T_\rho m_\rho^2
\int_0^1 du
h_{\parallel}^{(s)}(u) \frac{\Gamma(2)}{(-Q^2)^{2}}\nonumber \\
&&+\frac{1}{12}\langle\frac{ \alpha_sGG}{\pi}\rangle f_\rho m_\rho^3
\int_0^1 du \int^u_0 dt \int_0^t d\lambda C(\lambda) \frac{\Gamma(
2)}{(-Q^2)^{ 2}}\, ,
\end{eqnarray}

\begin{eqnarray}
\sqrt{2}\Pi_N^2&=&\frac{1}{\pi^2}f_\rho m_\rho^3 \int_0^1 du
\int^u_0 dt \int_0^t d\lambda
C(\lambda)\frac{\Gamma(1)}{(-Q^2)^{1}}\nonumber \\
&&+\frac{4}{3}\langle
\bar{q}q\rangle f^T_\rho m_\rho^2 \int_0^1 du h_{\parallel}^{(s)}(u) \frac{\Gamma(2)}{(-Q^2)^{2}}\nonumber \\
&&-\frac{1}{3}\langle \bar{q}g_s\sigma Gq\rangle f^T_\rho m_\rho^2
\int_0^1 du
h_{\parallel}^{(s)}(u) \frac{\Gamma(3)}{(-Q^2)^{3}}\nonumber \\
&&+\frac{1}{6}\langle\frac{ \alpha_sGG}{\pi}\rangle f_\rho m_\rho^3
\int_0^1 du \int^u_0 dt \int_0^t d\lambda
C(\lambda)\frac{\Gamma(3)}{(-Q^2)^{3}} \, ,
\end{eqnarray}

\begin{eqnarray}
\sqrt{2}\Pi_\Sigma^1&=&-\frac{1}{12\pi^2} f_\rho
m_\rho \int_0^1 du \left[\phi_{\parallel}(u)+6g_{\perp}^{(v)}(u)\right]\frac{\Gamma(\epsilon-1)}{(-Q^2)^{\epsilon-1}} \nonumber \\
&& +\frac{1}{\pi^2}f_\rho m_\rho^3 \int_0^1 du \int^u_0 dt \int_0^t
d\lambda C(\lambda) \frac{\Gamma(\epsilon)}{(-Q^2)^{\epsilon}}
\nonumber \\
&&+\frac{1}{16\pi^2} f_\rho
m_\rho^3 \int_0^1 du A(u)\frac{\Gamma(\epsilon)}{(-Q^2)^{\epsilon}}\nonumber\\
&&-\frac{1}{3}m_s \langle \bar{s}s\rangle f_\rho m_\rho \int_0^1 du
\left[\phi_{\parallel}(u) +2g_{\perp}^{(v)}(u)\right]\frac{\Gamma(1)}{(-Q^2)^{1}} \nonumber \\
&&+\frac{2}{3}\langle \bar{q}q\rangle f^T_\rho m_\rho^2 \int_0^1 du
h_{\parallel}^{(s)}(u)\frac{\Gamma(1)}{(-Q^2)^{1}}\nonumber\\
&&-\frac{1}{72} \langle\frac{ \alpha_sGG}{\pi}\rangle f_\rho
m_\rho \int_0^1 du \left[4g_{\perp}^{(v)}(u)+\phi_{\parallel}(u)\right] \frac{\Gamma(1)}{(-Q^2)^{1}}\nonumber \\
&& +\frac{1}{18} m_s \langle \bar{s}g_s\sigma Gs\rangle f_\rho
m_\rho \int_0^1 du
\phi_{\parallel}(u) \frac{\Gamma(2)}{(-Q^2)^{2}}\nonumber \\
&& +\frac{1}{12} m_s \langle \bar{s}s\rangle f_\rho m_\rho^3
\int_0^1 du A(u)\frac{\Gamma(2)}{(-Q^2)^{2}}
 \nonumber \\
&&  -\frac{1}{6}\langle \bar{q}g_s\sigma Gq\rangle f^T_\rho m_\rho^2
\int_0^1 du
h_{\parallel}^{(s)}(u)\frac{\Gamma(2)}{(-Q^2)^{2}}\nonumber \\
&&+\frac{4}{3} m_s \langle \bar{s}s\rangle f_\rho m_\rho^3\int_0^1
du \int^u_0 dt \int_0^t d\lambda
C(\lambda)\frac{\Gamma(2)}{(-Q^2)^{2}}\nonumber \\
 &&+\frac{1}{12} \langle\frac{
\alpha_sGG}{\pi}\rangle f_\rho m_\rho^3 \int_0^1 du \int^u_0 dt
\int_0^t d\lambda
C(\lambda) \frac{\Gamma(2)}{(-Q^2)^{2}}\nonumber \\
&&+\frac{1}{288}\langle\frac{ \alpha_sGG}{\pi}\rangle f_\rho
m_\rho^3 \int_0^1 du A(u)\frac{\Gamma(2)}{(-Q^2)^{2}} \, ,
\end{eqnarray}

\begin{eqnarray}
\sqrt{2}\Pi_\Sigma^2&=&-\frac{1}{6\pi^2}f_\rho m_\rho \int_0^1 du
\phi_{\parallel}(u)\frac{\Gamma(\epsilon)}{(-Q^2)^{\epsilon}}
\nonumber \\
&&+\frac{1}{8\pi^2}f_\rho
m_\rho^3 \int_0^1 du A(u)\frac{\Gamma(1)}{(-Q^2)^{1}}\nonumber \\
&& +\frac{2}{\pi^2}f_\rho m_\rho^3 \int_0^1 du \int^u_0 dt \int_0^t
d\lambda C(\lambda)\frac{\Gamma(1)}{(-Q^2)^{1}}
\nonumber\\
&&-\frac{2}{3}m_s \langle \bar{s}s\rangle f_\rho m_\rho \int_0^1 du
\phi_{\parallel}(u)\frac{\Gamma(2)}{(-Q^2)^{2}}
\nonumber \\
&&+\frac{4}{3}\langle \bar{q}q\rangle f^T_\rho m_\rho^2 \int_0^1 du
h_{\parallel}^{(s)}(u)\frac{\Gamma(2)}{(-Q^2)^{2}}\nonumber\\
&&-\frac{1}{36}\langle\frac{ \alpha_sGG}{\pi}\rangle f_\rho m_\rho
\int_0^1 du \phi_{\parallel}(u)\frac{\Gamma(2)}{(-Q^2)^{2}} \nonumber \\
&&+\frac{1}{9} m_s \langle \bar{s}g_s\sigma Gs\rangle
f_\rho m_\rho \int_0^1 du \phi_{\parallel}(u)\frac{\Gamma(3)}{(-Q^2)^{3}}\nonumber\\
&& +\frac{1}{6} m_s \langle \bar{s}s\rangle f_\rho m_\rho^3 \int_0^1
du A(u)\frac{\Gamma(3)}{(-Q^2)^{3}}
  \nonumber \\
&&-\frac{1}{3}\langle \bar{q}g_s\sigma Gq\rangle f^T_\rho m_\rho^2
\int_0^1 du
h_{\parallel}^{(s)}(u)\frac{\Gamma(3)}{(-Q^2)^{3}}\nonumber \\
&&+\frac{8}{3} m_s \langle \bar{s}s\rangle f_\rho m_\rho^3\int_0^1
du \int^u_0 dt \int_0^t d\lambda
C(\lambda)\frac{\Gamma(3)}{(-Q^2)^{3}}\nonumber \\
 &&+\frac{1}{6}\langle\frac{
\alpha_sGG}{\pi}\rangle f_\rho m_\rho^3 \int_0^1 du \int^u_0 dt
\int_0^t d\lambda
C(\lambda)\frac{\Gamma(3)}{(-Q^2)^{3}} \nonumber \\
&&+\frac{1}{144}\langle\frac{ \alpha_sGG}{\pi}\rangle f_\rho
m_\rho^3 \int_0^1 du A(u)\frac{\Gamma(3)}{(-Q^2)^{3}} \, ,
\end{eqnarray}

\begin{eqnarray}
\sqrt{2}\Pi_\Xi^1&=&-\frac{1}{12\pi^2}f_\rho m_\rho \int_0^1 du
\phi_{\parallel}(u)\frac{\Gamma(\epsilon-1)}{(-Q^2)^{\epsilon-1}}\nonumber \\
&&+\frac{1}{16\pi^2}f_\rho
m_\rho^3 \int_0^1 du A(u) \frac{\Gamma(\epsilon)}{(-Q^2)^{\epsilon}}\nonumber \\
&& +\frac{1}{2\pi^2}f_\rho m_\rho^3 \int_0^1 du \int^u_0 dt \int_0^t
d\lambda C(\lambda)\frac{\Gamma(\epsilon)}{(-Q^2)^{\epsilon}}
\nonumber\\
&&-\frac{2}{3}m_s \langle \bar{s}s\rangle f_\rho m_\rho \int_0^1 du
\left[\phi_{\parallel}(u) -2g_{\perp}^{(v)}(u)\right]\frac{\Gamma(1)}{(-Q^2)^{1}}  \nonumber \\
&&+\frac{1}{72}\langle\frac{ \alpha_sGG}{\pi}\rangle f_\rho
m_\rho \int_0^1 du \left[2g_{\perp}^{(v)}(u)-\phi_{\parallel}(u)\right]\frac{\Gamma(1)}{(-Q^2)^{1}} \nonumber \\
&& +\frac{1}{9} m_s \langle \bar{s}g_s\sigma Gs\rangle f_\rho m_\rho
\int_0^1 du
\phi_{\parallel}(u) \frac{\Gamma(2)}{(-Q^2)^{2}}\nonumber \\
&&+\frac{1}{6} m_s \langle \bar{s}s\rangle f_\rho m_\rho^3 \int_0^1
du A(u)\frac{\Gamma(2)}{(-Q^2)^{2}}
 \nonumber \\
 &&+\frac{1}{288}\langle\frac{
\alpha_sGG}{\pi}\rangle f_\rho m_\rho^3 \int_0^1 du A(u)
\frac{\Gamma(2)}{(-Q^2)^{2}}\, ,
\end{eqnarray}

\begin{eqnarray}
\sqrt{2}\Pi_\Xi^2&=&-\frac{1}{6\pi^2}f_\rho m_\rho \int_0^1 du
\phi_{\parallel}(u)\frac{\Gamma(\epsilon)}{(-Q^2)^{\epsilon}}
\nonumber \\
&&+\frac{1}{8\pi^2}f_\rho
m_\rho^3 \int_0^1 du A(u)\frac{\Gamma(1)}{(-Q^2)^{1}}\nonumber \\
&& +\frac{1}{\pi^2}f_\rho m_\rho^3 \int_0^1 du \int^u_0 dt \int_0^t
d\lambda C(\lambda)\frac{\Gamma(1)}{(-Q^2)^{1}}
\nonumber\\
&&-\frac{4}{3}m_s \langle \bar{s}s\rangle f_\rho m_\rho \int_0^1 du
\phi_{\parallel}(u)\frac{\Gamma(2)}{(-Q^2)^{2}}\nonumber \\
&&-\frac{1}{36}\langle\frac{ \alpha_sGG}{\pi}\rangle f_\rho m_\rho
\int_0^1 du \phi_{\parallel}(u)\frac{\Gamma(2)}{(-Q^2)^{2}}
\nonumber\\
&& +\frac{2}{9} m_s \langle \bar{s}g_s\sigma Gs\rangle
f_\rho m_\rho \int_0^1 du \phi_{\parallel}(u)\frac{\Gamma(3)}{(-Q^2)^{3}}\nonumber\\
&& +\frac{1}{3} m_s \langle \bar{s}s\rangle f_\rho m_\rho^3 \int_0^1
du A(u)\frac{\Gamma(3)}{(-Q^2)^{3}}
  \nonumber \\
&&+\frac{1}{144}\langle\frac{ \alpha_sGG}{\pi}\rangle f_\rho
m_\rho^3 \int_0^1 du A(u)\frac{\Gamma(3)}{(-Q^2)^{3}} \, ,
\end{eqnarray}
where $Q_\mu=q_\mu+up_\mu$, the $\epsilon$ is a small positive
quantity, after taking the double Borel transformation, we can take
the limit $\epsilon\rightarrow 0$. The light-cone distribution
amplitudes $\phi_{\parallel}(u)$, $g_{\perp}^{(v)}(u)$,
$h_{\parallel}^{(s)}(u)$, $C(u)$ and $A(u)$ of the $\rho$ meson are
presented in the appendix \cite{VMLC}, the non-perturbative
parameters in the light-cone distribution amplitudes are scale
dependent, in this article, the energy scale is taken to be
$\mu=1GeV$. Here we have neglected the contributions from the gluons
$G_{\mu \nu }$, the contributions proportional to the $G_{\mu\nu}$
can give rise to three-particle (and four-particle) meson
distribution amplitudes with a gluon (and quark-antiquark pair) in
addition to the two valence quarks, their corrections are usually
not expected to play any significant roles\footnote{For examples, in
the decay $B \to \chi_{c0}K$, the factorizable contribution is zero
and the non-factorizable contributions from the soft hadronic matrix
elements are too small to accommodate the experimental data
\cite{WangLH}; the net contributions from the three-valence particle
light-cone distribution amplitudes to the strong coupling constant
$g_{D_{s1}D^*K}$ are rather small, about $20\%$ \cite{Wang0611}. The
contributions of the three-particle (quark-antiquark-gluon)
distribution amplitudes of the mesons are always of minor importance
comparing with the two-particle (quark-antiquark) distribution
amplitudes in the light-cone QCD sum rules.   In our previous work,
we study the four form-factors $f_1(Q^2)$, $f_2(Q^2)$, $g_1(Q^2)$
and $g_2(Q^2)$ of the $\Sigma \to n$ in the framework of the
light-cone QCD sum rules approach up to twist-6 three-quark
light-cone distribution amplitudes and obtain satisfactory results
\cite{Wang06}. In the light-cone QCD sum rules,
 we can neglect the contributions from the valence gluons and make relatively rough estimations.}.

Matching the hadronic representations in Eq.(8) with the
corresponding ones in Eqs.(19-24) below the threshold $s^0_i$, then
perform the double Borel transformation with respect to the
variables $Q_1=-q^2$ and $Q_2=-(p+q)^2$ respectively, subtract the
contributions from the continuum states,
\begin{eqnarray}
B_{M_1}B_{M_2}
\frac{\Gamma[n]}{\left[u(1-u)m_\rho^2+(1-u)Q_1^2+uQ_2^2\right]^n}
 &=&\frac{M^{2(2-n)}}{M_1^2 M_2^2} e^{-\frac{u(1-u)m_\rho^2}{M^2}}
\delta(u-u_0) \, , \nonumber\\
\frac{1}{M^2}&=& \frac{1}{M^2_1}+\frac{1}{M^2_2} \, ,\nonumber\\
u_0&=& \frac{M_1^2}{M_1^2+M_2^2}\, ,\nonumber\\
 M^{2n}&\rightarrow& \frac{1}{\Gamma[n]}\int_0^{s_0} ds
s^{n-1}e^{-\frac{s}{M^2}} \, ,
\end{eqnarray}
finally we obtain the following six sum rules for the strong
coupling constants,
\begin{eqnarray}
g_N^V&=&-\frac{1}{2\sqrt{2}\lambda_N^2}e^{\frac{m_N^2-u_0(1-u_0)m_\rho^2}{M^2}}\left\{
\frac{1}{\pi^2}M^2E_0(x)f_\rho m_\rho^3 \int^{u_0}_0 dt \int_0^t
d\lambda C(\lambda) \right. \nonumber \\
&&+\frac{4}{3}\langle \bar{q}q\rangle f^T_\rho m_\rho^2
h_{\parallel}^{(s)}(u_0)-\frac{1}{3M^2}\langle \bar{q}g_s\sigma
Gq\rangle f^T_\rho m_\rho^2 h_{\parallel}^{(s)}(u_0) \nonumber\\
&&\left. +\frac{1}{6M^2}\langle\frac{ \alpha_sGG}{\pi} \rangle
f_\rho m_\rho^3 \int^{u_0}_0 dt \int_0^t d\lambda C(\lambda)
\right\} \, ,
\end{eqnarray}

 \begin{eqnarray}
g_\Sigma^V&=&-\frac{1}{2\sqrt{2}\lambda_\Sigma^2}e^{\frac{m_\Sigma^2-u_0(1-u_0)m_\rho^2}{M^2}}\left\{-\frac{1}{6\pi^2}M^4E_1(x)f_\rho
m_\rho \phi_{\parallel}(u_0) \right.\nonumber \\
&& +\frac{1}{8\pi^2}M^2E_0(x)f_\rho m_\rho^3
A(u_0)\nonumber\\
&&+\frac{2}{\pi^2}M^2E_0(x)f_\rho m_\rho^3   \int^{u_0}_0 dt
\int_0^t d\lambda C(\lambda)  \nonumber\\
&&-\frac{2}{3}m_s \langle \bar{s}s\rangle f_\rho m_\rho
\phi_{\parallel}(u_0)+\frac{4}{3}\langle \bar{q}q\rangle f^T_\rho
m_\rho^2 h_{\parallel}^{(s)}(u_0)\nonumber\\
&&-\frac{1}{36}\langle\frac{ \alpha_sGG}{\pi}\rangle f_\rho m_\rho
 \phi_{\parallel}(u_0) \nonumber\\
&& +\frac{1}{9M^2}  m_s \langle \bar{s}g_s\sigma Gs\rangle f_\rho
m_\rho   \phi_{\parallel}(u_0)\nonumber\\
&&+\frac{1}{6M^2}  m_s \langle \bar{s}s\rangle f_\rho m_\rho^3
  A(u_0)  \nonumber \\
&&-\frac{1}{3M^2} \langle \bar{q}g_s\sigma Gq\rangle f^T_\rho
m_\rho^2 h_{\parallel}^{(s)}(u_0)\nonumber\\
&&+\frac{8}{3M^2}  m_s \langle \bar{s}s\rangle f_\rho m_\rho^3
\int^{u_0}_0 dt \int_0^t d\lambda
C(\lambda)\nonumber \\
 && +\frac{1}{6M^2}\langle\frac{ \alpha_sGG}{\pi}\rangle
f_\rho m_\rho^3   \int^{u_0}_0 dt \int_0^t d\lambda
C(\lambda)\nonumber\\
&&\left.+\frac{1}{144M^2}\langle\frac{ \alpha_sGG}{\pi}\rangle
f_\rho m_\rho^3    A(u_0) \right\} \, ,
\end{eqnarray}

\begin{eqnarray}
g_\Xi^V&=&-\frac{1}{2\sqrt{2}\lambda_\Xi^2}e^{\frac{m_\Xi^2-u_0(1-u_0)m_\rho^2}{M^2}}\left\{-\frac{1}{6\pi^2}M^4E_1(x)f_\rho
m_\rho \phi_{\parallel}(u_0) \right.\nonumber \\
&& +\frac{1}{8\pi^2}M^2E_0(x)f_\rho m_\rho^3
A(u_0)\nonumber\\
&&+\frac{1}{\pi^2}M^2E_0(x)f_\rho m_\rho^3   \int^{u_0}_0 dt
\int_0^t d\lambda C(\lambda)
\nonumber\\
&&-\frac{4}{3} m_s \langle \bar{s}s\rangle f_\rho m_\rho
\phi_{\parallel}(u_0)-\frac{1}{36}\langle\frac{
\alpha_sGG}{\pi}\rangle f_\rho m_\rho  \phi_{\parallel}(u_0)
\nonumber\\
&& +\frac{2}{9M^2}  m_s \langle \bar{s}g_s\sigma Gs\rangle f_\rho
m_\rho   \phi_{\parallel}(u_0) \nonumber\\
&&+\frac{1}{3M^2}  m_s \langle \bar{s}s\rangle f_\rho m_\rho^3
 A(u_0)\nonumber\\
&& \left. +\frac{1}{144M^2}\langle\frac{ \alpha_sGG}{\pi}\rangle
f_\rho m_\rho^3 A(u_0) \right\} \, ,
\end{eqnarray}

\begin{eqnarray}
g^V_N+g^T_N&=&-\frac{\sqrt{2}}{\lambda_N^2
m_\rho^2}e^{\frac{m_N^2-u_0(1-u_0)m_\rho^2}{M^2}}\left\{-\frac{1}{2\pi^2}M^6E_2(x)f_\rho
m_\rho g_{\perp}^{(v)}(u_0)  \right.
\nonumber\\
&& +\frac{1}{2\pi^2}M^4E_1(x)f_\rho m_\rho^3
  \int^{u_0}_0 dt \int_0^t d\lambda C(\lambda) \nonumber\\
 && +\frac{2}{3}M^2E_0(x)\langle \bar{q}q\rangle f^T_\rho m_\rho^2
h_{\parallel}^{(s)}(u_0)\nonumber\\
&&-\frac{1}{12}M^2E_0(x)\langle\frac{ \alpha_sGG}{\pi}\rangle f_\rho
m_\rho  g_{\perp}^{(v)}(u_0) \nonumber \\
&&-\frac{1}{6}\langle \bar{q}g_s\sigma Gq\rangle f^T_\rho m_\rho^2
 h_{\parallel}^{(s)}(u_0) \nonumber\\
&&\left.+\frac{1}{12}\langle\frac{ \alpha_sGG}{\pi} \rangle f_\rho
m_\rho^3  \int^{u_0}_0 dt \int_0^t d\lambda C(\lambda) \right\} \,,
\end{eqnarray}

\begin{eqnarray}
g^V_\Sigma+g^T_\Sigma&=&-\frac{\sqrt{2}}{\lambda_\Sigma^2
m_\rho^2}e^{\frac{m_\Sigma^2-u_0(1-u_0)m_\rho^2}{M^2}}\left\{
\right.\nonumber \\
&&-\frac{1}{12\pi^2}M^6 E_2(x)f_\rho m_\rho
\left[\phi_{\parallel}(u_0)+6g_{\perp}^{(v)}(u_0)\right]\nonumber\\
&& +\frac{1}{16\pi^2}M^4 E_1(x)f_\rho m_\rho^3
A(u_0)\nonumber\\
&&+\frac{1}{\pi^2}M^4 E_1(x)f_\rho m_\rho^3  \int^{u_0}_0 dt
\int_0^t d\lambda C(\lambda) \nonumber\\
&&-\frac{1}{3}M^2E_0(x)m_s \langle \bar{s}s\rangle f_\rho m_\rho
\left[\phi_{\parallel}(u_0) +2g_{\perp}^{(v)}(u_0)\right] \nonumber \\
&&+\frac{2}{3}M^2E_0(x)\langle \bar{q}q\rangle f^T_\rho m_\rho^2
h_{\parallel}^{(s)}(u_0)\nonumber\\
&&-\frac{1}{72}M^2E_0(x) \langle\frac{ \alpha_sGG}{\pi}\rangle
f_\rho
m_\rho   \left[4g_{\perp}^{(v)}(u_0)+\phi_{\parallel}(u_0)\right] \nonumber \\
&& +\frac{1}{18} m_s \langle \bar{s}g_s\sigma Gs\rangle f_\rho
m_\rho   \phi_{\parallel}(u_0) +\frac{1}{12} m_s \langle
\bar{s}s\rangle f_\rho m_\rho^3   A(u_0)
 \nonumber \\
&&  -\frac{1}{6}\langle \bar{q}g_s\sigma Gq\rangle f^T_\rho m_\rho^2
  h_{\parallel}^{(s)}(u_0)\nonumber\\
&&+\frac{4}{3} m_s \langle \bar{s}s\rangle f_\rho m_\rho^3
\int^{u_0}_0 dt \int_0^t d\lambda
C(\lambda)\nonumber \\
 &&+\frac{1}{12} \langle\frac{
\alpha_sGG}{\pi}\rangle f_\rho m_\rho^3   \int^{u_0}_0 dt \int_0^t
d\lambda C(\lambda) \nonumber\\
&&\left.+\frac{1}{288}\langle\frac{ \alpha_sGG}{\pi}\rangle f_\rho
m_\rho^3
 A(u_0) \right\} \, ,
\end{eqnarray}

\begin{eqnarray}
g^V_\Xi+g^T_\Xi&=&-\frac{\sqrt{2}}{\lambda_\Xi^2
m_\rho^2}e^{\frac{m_\Xi^2-u_0(1-u_0)m_\rho^2}{M^2}}\left\{-\frac{1}{12\pi^2}M^6E_2(x)f_\rho
m_\rho \phi_{\parallel}(u_0)\right.\nonumber \\
&& +\frac{1}{16\pi^2}M^4E_1(x)f_\rho m_\rho^3   A(u_0)
\nonumber \\
&&+\frac{1}{2\pi^2}M^4E_1(x)f_\rho m_\rho^3   \int^{u_0}_0 dt
\int_0^t
d\lambda C(\lambda) \nonumber \\
&&-\frac{2}{3}M^2E_0(x)m_s \langle \bar{s}s\rangle f_\rho m_\rho
\left[\phi_{\parallel}(u_0) -2g_{\perp}^{(v)}(u_0)\right]  \nonumber \\
&&+\frac{1}{72}M^2E_0(x)\langle\frac{ \alpha_sGG}{\pi}\rangle f_\rho
m_\rho   \left[2g_{\perp}^{(v)}(u_0)-\phi_{\parallel}(u_0)\right] \nonumber \\
&& +\frac{1}{9} m_s \langle \bar{s}g_s\sigma Gs\rangle f_\rho m_\rho
  \phi_{\parallel}(u_0) +\frac{1}{6} m_s \langle
\bar{s}s\rangle f_\rho m_\rho^3   A(u_0)\nonumber \\
&& \left.+\frac{1}{288}\langle\frac{ \alpha_sGG}{\pi}\rangle f_\rho
m_\rho^3
 A(u_0) \right\} \, ,
\end{eqnarray}
where
\begin{eqnarray}
E_n(x)&=&1-(1+x+\frac{x^2}{2!}+\cdots+\frac{x^n}{n!})e^{-x} \, , \nonumber\\
x&=&\frac{s^0_i}{M^2} \, ,\nonumber
\end{eqnarray}
the $s^0_i$ are the threshold parameters.

\section{Numerical result and discussion}
The input parameters are taken as $\langle \bar{q}q
\rangle=-(0.24\pm 0.01 GeV)^3$, $\langle \bar{s}s \rangle=(0.8\pm
0.1 )\langle \bar{q}q \rangle$, $\langle \bar{q}g\sigma Gq
\rangle=m_0^2\langle \bar{q}q \rangle$, $\langle \bar{s}g\sigma Gs
\rangle=m_0^2\langle \bar{s}s \rangle$, $m_0^2=(0.8 \pm 0.1)GeV^2$,
$\langle \frac{\alpha GG}{\pi}\rangle=(0.33GeV)^4 $, $m_u=0$,
$m_d=0$, $m_s=(140\pm 10 )MeV$ \cite{LCSR,LCSRreview,SVZ79},
$f_\rho=(0.198\pm0.007)GeV$, $f_\rho^T=(0.160\pm0.010)GeV$,
 $a_2^{\perp}=0.20\pm0.10$,
$a_2^{\parallel}=0.18\pm0.10$, $\varsigma_3=0.032\pm 0.010$,
$\varsigma_4=0.15\pm 0.10$, $\varsigma_4^T=0.10\pm 0.05$,
$\widetilde{\varsigma}_4^T=-0.10\pm 0.05$, $\omega_3^A=-2.1\pm1.0$,
$\omega_3^V=3.8\pm1.8$, $\omega_3^T=7.0\pm7.0$ \cite{VMLC},
 $m_\rho=0.77GeV$, $m_{p}=0.938GeV$, $m_{\Sigma}=1.189GeV$,
$m_{\Xi}=1.315GeV$, $\lambda_{p}=(2.4\pm 0.2)\times 10^{-2}GeV^3$,
$\lambda_{\Sigma}=(3.2\pm0.2)\times 10^{-2}GeV^3$,
$\lambda_{\Xi}=(3.8\pm 0.2)\times 10^{-2}GeV^3$, $s^0_{p}=2.3GeV^2$,
$s^0_{\Sigma}=3.2GeV^2$ and $s^0_{\Xi}=3.6GeV^2$ \cite{Pasupathy85}.
The values of the vacuum condensates have been updated with the
experimental data for the $\tau$ decays, the QCD sum rules for the
baryon masses and analysis of the charmonium spectrum
\cite{Zyablyuk, Ioffe2005,AlphaS},  in this article, we choose the
standard (or old) values to keep in consistent with the sum rules
used in determining the non-perturbative parameters in the
light-cone distribution amplitudes.  The Borel parameters are chosen
as $ M_1^2=M_2^2= (2-4)GeV^2$ and $M^2=(1-2)GeV^2$, in those
regions, the values of the strong coupling constants $g^V_N$,
$g^V_\Sigma$ and $g^V_\Xi$ are rather stable with the variation of
the Borel parameter $M^2$ from the sum rules in Eqs.(26-28), while
the sum rules for the $g^V_N+g^T_N$, $g^V_\Sigma+g^T_\Sigma$ and
$g^V_\Xi+g^T_\Xi$ in Eqs.(29-31) are not as stable as the
corresponding ones for the $g^V_N$, $g^V_\Sigma$ and $g^V_\Xi$,
which are shown in the Figs.(1-3).

Taking into account all the uncertainties, finally we obtain the
numerical results of the strong coupling constants $g^V_N$,
$g^V_\Sigma$, $g^V_\Xi$, $g^V_N+g^T_N$, $g^V_\Sigma+g^T_\Sigma$ and
$g^V_\Xi+g^T_\Xi$, which are shown in the Figs.1-3,
\begin{eqnarray}
  g^V_N &=&3.2 \pm0.9 \, ,\nonumber \\
  g^V_\Sigma &=&4.0 \pm 1.0 \, ,\nonumber \\
  g^V_\Xi &=&1.5 \pm 1.1 \, ,\nonumber \\
g^V_N+g^T_N &=&36.8 \pm 13.0 \, ,\nonumber \\
g^V_\Sigma+g^T_\Sigma &=&53.5 \pm 19.0 \, ,\nonumber \\
g^V_\Xi+g^T_\Xi &=&-5.3 \pm 3.3 \, .
\end{eqnarray}
The numerical values are in agreement with  the existing
calculations in part, for examples, the values of the external field
QCD sum rules ($g^V_N =2.4\pm0.6 $,
  $g^V_\Sigma =4.8\pm1.2$
  $g^V_\Xi =2.4\pm 0.6$,
$g^T_N =7.7\pm1.9$, $g^T_\Sigma =2.3\pm0.4$,
 $g^T_\Xi
=-5.0\pm 1.0$) \cite{Rijken06} and the phenomenologically fitted
values of the one-boson exchange model ($g^V_N =3.0$,
  $g^V_\Sigma =5.9$
  $g^V_\Xi =3.0$,
$g^T_N =12.5$, $g^T_\Sigma =9.1$,
 $g^T_\Xi
=-3.4$) \cite{Nijmegen2}. In calculation, we observe that the main
contributions come from the perturbative terms and the quark
condensates terms ($\langle \bar{q}q\rangle$ and $\langle
\bar{s}s\rangle$), the contributions of the mixed condensates and
gluon condensates are of minor importance. It is not un-expected.
From the Table.1, we can see that the contributions from the terms
of quark condensates  are comparable with  the perturbative terms in
the region $M^2=(1-2)GeV^2$. The presence of the chiral symmetry
breaking quark condensates and the existence of the massive baryons
are closely related to each other, $m_p\simeq -\frac{8\pi^2\langle
\bar{q}q\rangle}{M^2}$ \cite{Ioffe81}, the contributions from the
quark condensates (of   order $ \mathcal {O}(M^2)$ or $ \mathcal
{O}(M^0)$) in the sum rules in Eqs.(26-31) are of great importance,
the subtractions of the continuum states can be implemented by the
simple replacement $M^{2n}\rightarrow
 M^{2n} E_{n-1}(x)$ for $n\geq 1$, we have to introduce some unknown parameters $C_A$ and $C_B$ to
 subtract the contributions from the continuum states  of order $\mathcal {O}(M^0)$
 with  $g_V\rightarrow g_V+C_A$ and  $g_V+g_T\rightarrow g_V+g_T+C_B$, however, it is difficult  to obtain  the
values of the $C_A$ and $C_B$. In the region $M^2=(1-2)GeV^2$,
$\frac{\alpha_s(M)}{\pi}\sim 0.14-0.18$ \cite{AlphaS}, if the
radiative $\alpha_s$ corrections to the perturbative terms are
companied  with large numerical factors, just like in the case of
the QCD sum rules for the masses of the proton \cite{Ioffe2005}, the
contributions of  order $\mathcal {O}(\alpha_s)$  may be large,
neglecting them can impair the predictive power.  The contributions
of the three-particle (quark-quark-gluon) light-cone distribution
amplitudes  may be of the same order as the mixed condensates
($\langle \bar{q}g_s \sigma Gq\rangle$ and $\langle \bar{s}g_s
\sigma Gs\rangle$), neglecting them would not impair the predictive
power much. The consistent and complete light-cone QCD sum rules
analysis should include the contributions from the perturbative
$\alpha_s$ corrections, the distribution amplitudes with additional
valence gluons and quark-antiquark pairs, the reasonable
subtractions of the continuum states of   order $\mathcal {O}(M^0)$,
and improve the parameters which enter in the light-cone QCD sum
rules, that may be our next work.

\begin{table}[htb]
\begin{center}
\begin{tabular}{c|c|c|c|c|c|c} \hline\hline &$g_N^V$&$g_N^V+g_N^T$&$g_\Sigma^V$&$g_\Sigma^V+g_\Sigma^T$&
$g_\Xi^V$&$g_\Xi^V+g_\Xi^T$  \\\hline
  $PT$&$ \mathcal {O}(M^2)$ &$ \mathcal {O}(M^6,M^4)$&$\mathcal {O}(M^4,M^2)$&$\mathcal {O}(M^6,M^4)$&$ \mathcal {O}(M^4,M^2)$ &$\mathcal {O}(M^6,M^4)$  \\\hline
  $\langle\bar{q}q\rangle$&$ \mathcal {O}(M^0)$ &$ \mathcal {O}(M^2)$&$\mathcal {O}(M^0)$&$\mathcal {O}(M^2,M^0)$&$\mathcal {O}(M^0)$ &$\mathcal {O}(M^2,M^0)$  \\\hline\hline
\end{tabular}
\end{center}
\caption{The power counting order of the   $g^V_N$, $g^V_\Sigma$,
  $g^V_\Xi$, $g^V_N+g^T_N $, $g^V_\Sigma+g^T_\Sigma $,
$g^V_\Xi+g^T_\Xi$, respectively. The $PT$ stands for the
perturbative terms and the $\langle\bar{q}q\rangle$ stands for
contributions from the quark condensates, we have not shown the
terms which are suppressed by negative powers of the $M^2$
explicitly. }
\end{table}

\begin{figure}
\centering
  \includegraphics[totalheight=7cm,width=7cm]{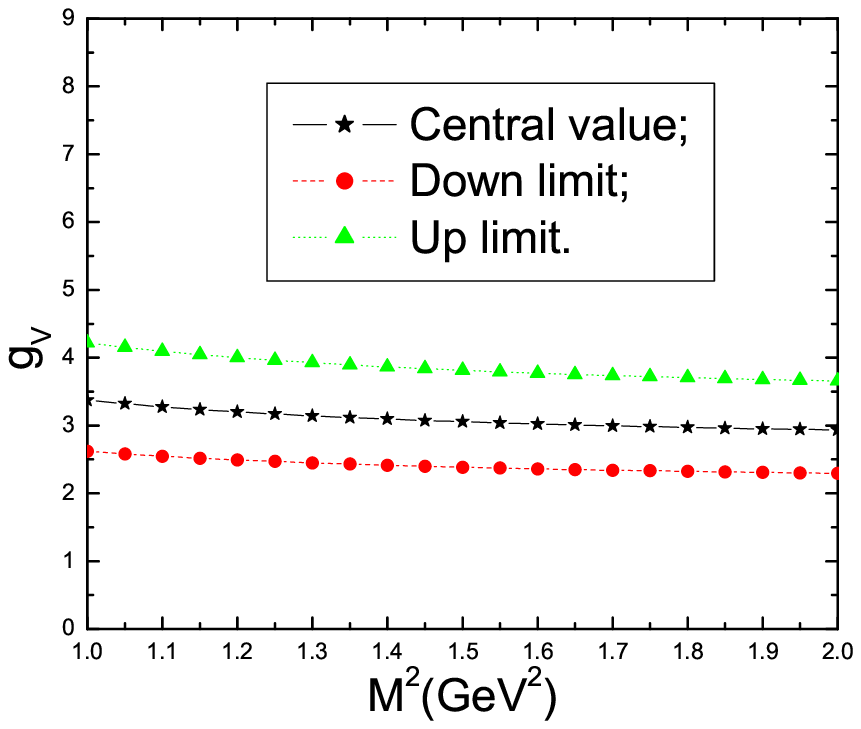}
  \includegraphics[totalheight=7cm,width=7cm]{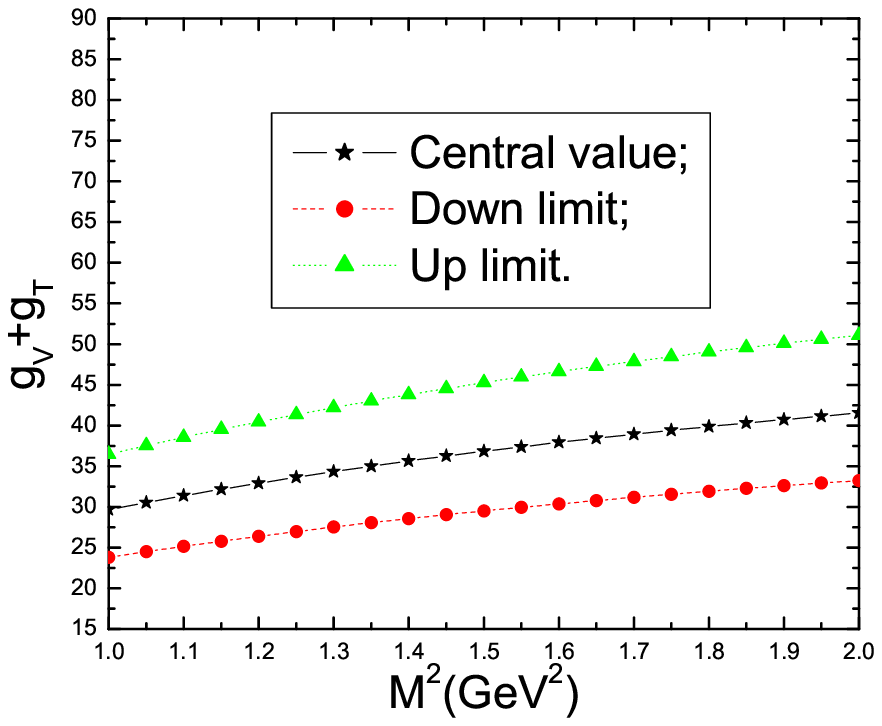}
   \caption{The   $g^V_N$ and $g^V_N+g^T_N$ with the Borel parameter $M^2$ . }
\end{figure}

\begin{figure}
\centering
  \includegraphics[totalheight=7cm,width=7cm]{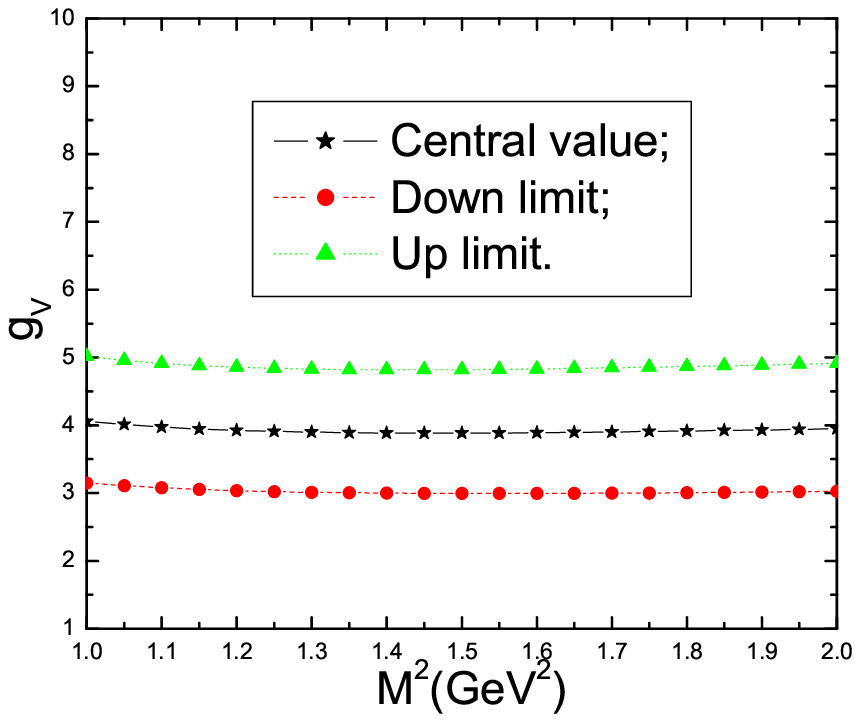}
  \includegraphics[totalheight=7cm,width=7cm]{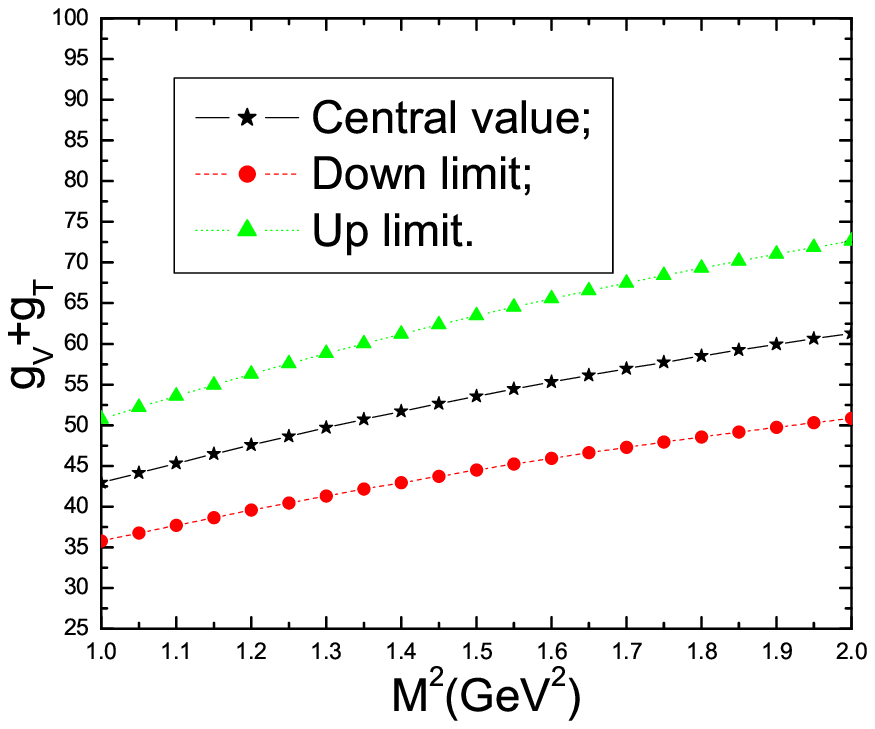}
   \caption{The   $g^V_\Sigma$ and $g^V_\Sigma+g^T_\Sigma$ with the Borel parameter $M^2$. }
\end{figure}

\begin{figure}
\centering
  \includegraphics[totalheight=7cm,width=7cm]{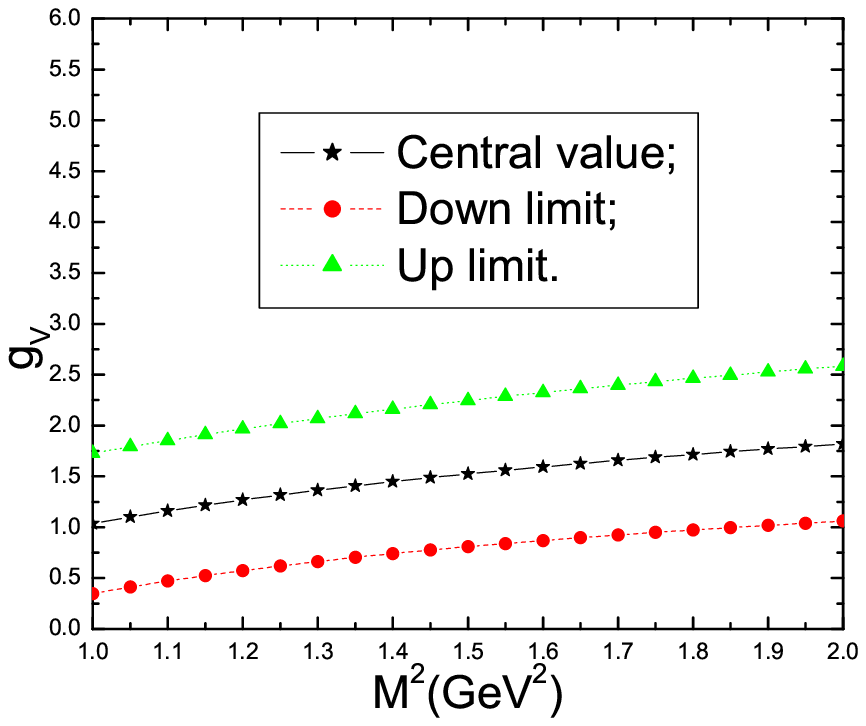}
  \includegraphics[totalheight=7cm,width=7cm]{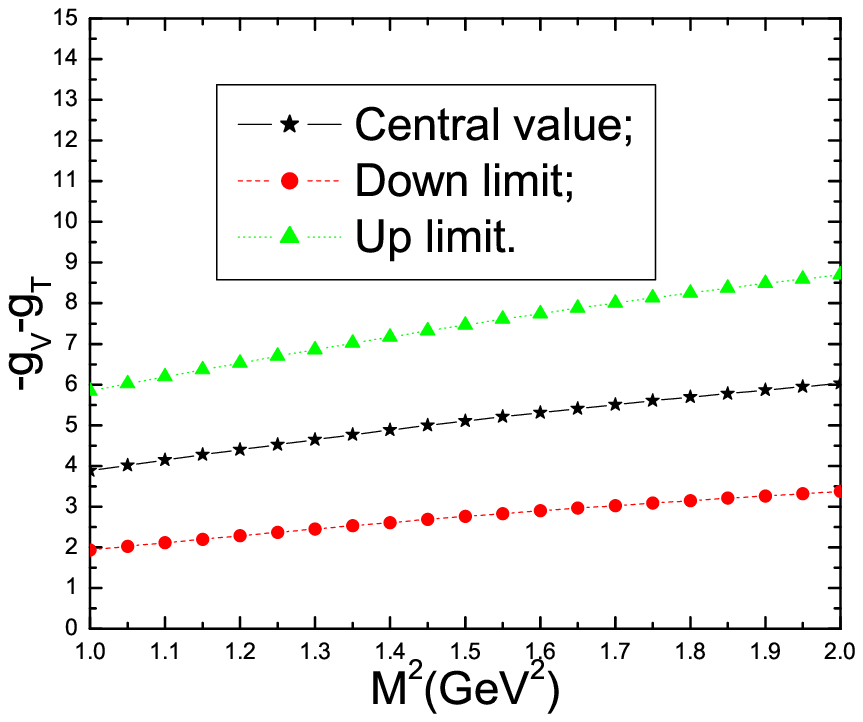}
   \caption{The   $g^V_\Xi$ and $g^V_\Xi+g^T_\Xi$ with the Borel parameter $M^2$. }
\end{figure}

The  Clebsch-Gordan coefficients of the octet VNN vertices indicate
that \cite{CGcoeff}
\begin{eqnarray}
g_\Sigma&=&2\alpha g_N \, , \nonumber \\
 g_\Xi &=&(2\alpha-1) g_N \, ,
\end{eqnarray}
where we use the same notation $\alpha$ to present the electric and
magnetic ratios $\alpha_e$ and $\alpha_m$. Our numerical values
$\alpha_e\sim 0.6-0.7$ and $\alpha_m \sim 0.4-0.7$ deviate greatly
from the prediction of the vector meson dominance theory,
$\alpha_e=1$ \cite{VMD}. The numerical values of the $g^V_N$,
  $g^V_\Sigma$,  $g^V_\Xi$, $g^T_N$, $g^T_\Sigma$,  $g^T_\Xi$ do not
  obey the simple  relation obtained from the group theory in Eq.(33), the $SU(3)$ symmetry
  breaking effects are very large.

\section{Conclusion}
In this article, we calculate the strong coupling constants $g^V_N$,
  $g^V_\Sigma$,  $g^V_\Xi$, $g^T_N$, $g^T_\Sigma$,  $g^T_\Xi$ of
the $\rho NN$, $\rho\Sigma\Sigma$ and $\rho\Xi\Xi$  in the framework
of the light-cone QCD sum rules approach. The  strong coupling
constants of the meson-baryon-baryon are the fundamental parameters
in the one-boson exchange model which describes the baryon-baryon
interactions successfully. The numerical results are in agreement
with the existing calculations in part, for examples, the
predictions of the external field QCD sum rules and the
phenomenologically fitted values of the one-boson exchange model.
The electric and magnetic $F/(F+D)$ ratios deviate  greatly from the
prediction of the vector meson dominance theory, the $SU(3)$
symmetry breaking effects are very large. Those deviations may be
due to the  failure to take into account the perturbative $\alpha_s$
corrections, the un-subtracted contributions from   the continuum
states of  order $\mathcal {O}(M^0)$, and the neglected distribution
amplitudes with additional valence gluons and quark-antiquark pairs.
The consistent and complete light-cone QCD sum rules analysis should
include the contributions from the perturbative $\alpha_s$
corrections, the distribution amplitudes with additional valence
gluons and quark-antiquark pairs, the reasonable subtractions of
continuum states of  order $\mathcal {O}(M^0)$, and improve the
parameters which enter in the light-cone QCD sum rules.
 \section*{Appendix}
 The light-cone distribution amplitudes of the $\rho$ meson are defined
 by \cite{VMLC}
 \begin{eqnarray}
\langle 0| {\bar u} (x) \gamma_\mu d(0) |\rho(p)\rangle& =& p_\mu
f_\rho m_\rho \frac{\epsilon \cdot x}{p \cdot x} \int_0^1 du  e^{-i
u p\cdot x} \left\{\phi_{\parallel}(u)+\frac{m_\rho^2x^2}{16}
A(u)\right\}\nonumber\\
&&+\left[ \epsilon_\mu-p_\mu \frac{\epsilon \cdot x}{p \cdot x}
\right]f_\rho m_\rho
\int_0^1 du  e^{-i u p \cdot x} g_{\perp}^{(v)}(u)  \nonumber\\
&&-\frac{1}{2}x_\mu \frac{\epsilon \cdot x}{(p \cdot x)^2} f_\rho
m_\rho^3 \int_0^1 du e^{-iup \cdot x}C(u) \, ,
 \end{eqnarray}
 \begin{eqnarray}
 \langle 0| {\bar u} (x)  d(0) |\rho(p)\rangle  &=& -\frac{i}{
2}\left[f_\rho^T-f_\rho \frac{m_u+m_d}{m_\rho}\right]m_
\rho^2\epsilon \cdot x  \int_0^1 du  e^{-i u p \cdot x}
h_{\parallel}^{(s)}(u)  \, .
 \end{eqnarray}
The  light-cone distribution amplitudes are parameterized as
\cite{VMLC}
\begin{eqnarray}
\phi_{\parallel}(u,\mu)&=&6u(1-u) \left\{1+a_2^{\parallel}
\frac{3}{2}(5\xi^2-1) \right\}\, , \nonumber\\
g_{\perp}^{(v)}(u,\mu)&=&\frac{3}{4}(1+\xi^2)+\left\{ \frac{3}{7}a_2^{\parallel}+ 5\varsigma_3\right\}(3\xi^2-1)\nonumber \\
&&+\left\{\frac{9}{112}a_2^{\parallel}+\frac{15}{64}\varsigma_3(3\omega^V_3-\omega^A_3 )\right\}(3-30\xi^2+35\xi^4)\, ,  \nonumber \\
g_3(u,\mu)&=&1+\left\{ -1-\frac{2}{7}a_2^{\parallel}+\frac{40}{3}\varsigma_3 -\frac{20}{3}\varsigma_4\right\}C_2^{\frac{1}{2}}(\xi)\nonumber \\
&&+\left\{-\frac{27}{28}a_2^{\parallel} +\frac{5}{4}\varsigma_3 -\frac{15}{16}\varsigma_3(\omega^A_3+3\omega^V_3)\right\}C_4^{\frac{1}{2}}(\xi)\, ,  \nonumber \\
h_{\parallel}^{(s)}(u,\mu)&=&6u(1-u)
\left\{1+(\frac{1}{4}a_2^{\parallel}+\frac{5}{8}\varsigma_3\omega_3^T)
(5\xi^2-1) \right\}\, , \nonumber\\
A(u,\mu)&=&30u^2(1-u)^2\left\{\frac{4}{5}
+\frac{4}{105}a_2^{\parallel}+\frac{20}{9}\varsigma_4+\frac{8}{9}\varsigma_3\right\}\,
, \nonumber \\
C(u,\mu)&=&g_3(u,\mu)+\phi_{\parallel}(u,\mu)-2g_{\perp}^{(v)}(u,\mu)
\, ,
\end{eqnarray}
where the $\xi=2u-1$, and  the   $ C_2^{\frac{1}{2}}$ and
$C_4^{\frac{1}{2}}$
   are Gegenbauer polynomials \cite{VMLC}.
\section*{Acknowledgments}
This  work is supported by National Natural Science Foundation,
Grant Number 10405009,  and Key Program Foundation of NCEPU.

\end{document}